\documentclass[twocolumn,prl,showpacs,preprintnumbers,amsmath,amssymb]{revtex4}
\usepackage{graphicx}% Include figure files
\usepackage{dcolumn}% Align table columns on decimal point
\usepackage{amsmath}% bold math
\begin{document} 
\title{Oblate-prolate shape coexistence at finite angular momentum}
\author{Daniel Almehed}
\email{D.Almehed@umist.ac.uk}
\affiliation{Department of Physics, UMIST, P.O. Box 88, 
Manchester M60 1QD, United Kingdom} 
\author{Niels R.~Walet} 
\email{Niels.Walet@umist.ac.uk}
\affiliation{Department of Physics, UMIST, P.O. Box 88, 
Manchester M60 1QD, United Kingdom} 
 
\begin{abstract}
We investigate shape coexistence in a rotating nucleus.
We concentrate on the interesting
case of $^{72}$Kr which exhibits an interesting interplay between
prolate and oblate states as a function of angular momentum. The
calculation uses the local harmonic version of the method of
self-consistent adiabatic large-amplitude collective motion. We find
that the collective behaviour of the system changes with angular
momentum and we focus on the role of non-axial shapes.
\end{abstract} 
\maketitle 

Only a small number of quantum mechanical many-body systems can be
solved exactly through analytical or numerical means. Many approximate
schemes exist to find reasonable answers, but these may not always
be easy to interpret in physical terms. Another slant on this problem
is to try and describe some of the dynamics of such systems in terms of
a limited set of degrees of freedom.  These should of course be chosen
through some method appropriate for the problem at hand. Many
approaches are available, in areas ranging from field theories to
atomic physics (see, e.g., the reviews in Ref.~\cite{To98}). These are
typically based on the concept of ``relevant degrees of freedom'', or
on the introduction of collective motion and collective paths -- which
are two ways to express rather similar principles! Depending on the
energy scales involved and the physical situation being described, the
remaining degrees of freedom are either frozen out, or treated as a
heat-bath for the motion of the relevant degrees of freedom.

The most common description of nuclear collective dynamics is influenced
by the success of the liquid drop model~\cite{BK37}. The description is 
based on the competition between quadrupole shape fluctuations of the 
nucleus, typically modelled microscopically by a long-range
quadrupole-quadrupole force, and a BCS-like pair condensate, modelled 
by a short-range pairing force. The model of the nucleus does not have
to be this naive, but even with better models, it is often assumed
that the low-energy dynamics is well described by quadrupole shape
fluctuations. We would like to question this assumption, which leads
to a search for the best choice of collective coordinates (the
standard terminology for the relevant degrees of freedom in this
field). There are quite a few partial answers, see the
review~\cite{DK00} for a discussion of some of these. The goal is to
find a method that determines a collective path self-consistently,
based only on knowledge of the Hamiltonian governing the system. 

Clearly, in nuclear physics one does not know the Hamiltonian. However, 
for the collective properties of medium mass to heavy nuclei, models that
contain the two key parts of the nuclear force, a short-range pairing
force, and a long range multipole-multipole force (usually approximated
by a quadrupole-quadrupole one) are known to be able to capture the
essential part of the physics, see e.g. the textbook Ref.~\cite{RS}.

With modern experimental techniques, we can create nuclei at
considerable angular momentum, and study the behaviour as a function
of angular momentum, which provides us with an additional parameter we are
almost free to choose. The resulting experimental data  are
often  analysed in terms of simple collective models, usually
expressed in terms of quadrupole shapes.  The physics of collective
motion is then often described in terms of shape transitions and shape
coexistence, the mixing between various shapes.  One important example
is the recent large interest in shape coexistence at low angular
momentum in nuclei in the $A=70$ and $N \approx Z$
mass-region~\cite{BM03,FB00,PS00,YM01,KN03}. In this region
one finds a large diversity of shapes, and rapid changes in shape with
particle number and angular momentum. This large variety  is
caused by shell gaps corresponding to spherical ($N,Z=38$), prolate 
($N,Z=34,38,40$) and oblate ($N,Z=34,36$) shapes that exist in this
mass-region. The detailed understanding of how states with different
shapes mix is, among others, important for the astrophysical
nucleosynthesis process that passes through these proton-rich nuclei.

The quality of the recent experimental data, fuelled by advances within
the field of $\gamma$-ray spectroscopy, has led to theoretical effort
focused on this mass-region. The nucleus $^{72}$Kr, the subject of
this letter, shows oblate-prolate shape coexistence and/or shape
transitions, the nature of which seems to depend strongly on angular
momentum. There are other calculations for this nucleus: Ref.~\cite{YM01} 
have used the constrained mean-field method to study the potential energy 
surface. The approach used in Ref.~\cite{PS00} goes beyond the mean field 
approximation but does not answer the question which degrees of freedom 
are important for the collective path. Oblate-prolate shape coexistence
has also recently been studied in other mass regions, see e.g.~Ref.~\cite{BB04}. 

Our formalism, as set out in detail in~\cite{DK00,AW04}, is based on
time-dependent mean field theory, and the fact that a classical
dynamics can be associated with it. The issue of selecting collective
coordinates, and determining their coupling to other degrees of
freedom, thus becomes an exercise in classical mechanics.  If we
assume slow motion the mean-field energy, which is also the classical
Hamiltonian, can be expanded to second order. This corresponds to a
parametrisation of the one-body density matrix in terms of a set of
canonical coordinates, $\xi^\alpha$, and conjugate momenta,
$\pi_\alpha$. The potential $V(\xi)$ and the mass matrix $B^{\alpha
\beta}(\xi)$ are the coefficients of the expansion of the classical
Hamiltonian in powers of the momenta $\pi$ at zeroth and second
order, respectively.  Within this adiabatic Hamiltonian we search for
collective (and non-collective) coordinates $q^\mu$ and conjugate
momenta $p_\mu$. These are assumed to be obtained by an invertible
point transformation of the original coordinates $\xi^\alpha$ and
momenta $\pi_\alpha$, preserving the quadratic truncation of the
momentum dependence of the Hamiltonian, by
\begin{equation}
	\label{eq:fg}
	q^\mu = f^\mu(\xi), \quad p_\mu = g^\alpha_{,\mu}\pi_\alpha \quad \left( 
	\mu , \alpha =  1,\ldots ,n \right) 
\end{equation}
where we use a standard notation for the derivatives.
The adiabatic Hamiltonian is then transformed into
\begin{equation}
	\label{eq:H3}
	\bar{{\cal H}}(q,p) = \bar{V}(q) + \frac{1}{2} \bar{B}^{\mu \nu} 
	p_\mu p_\nu + {\cal O}(p^4) 
\end{equation}
in the new coordinates. The new coordinates $q^\mu$ are now to be divided
into three  categories: the collective coordinate, the
zero-mode coordinates, which describe motions that
do not change the energies and finally the remaining non-collective
coordinates. 

In the local harmonic approximation (LHA) the collective coordinate is 
determined by means of the solution to a set of self-consistent equations.  
These are the force equations and the local RPA equation
\begin{eqnarray}
	\label{eq:force1}
	\bar{{\cal H}}_{,\alpha} &=& \Lambda f_{,\alpha} +\omega J_{x,\alpha}+
	\sum_{\tau = n,p} \mu_\tau N_{\tau ,\alpha}
%+ \Lambda_I f^I_{,\alpha} 
,\\
	\label{eq:localRPA1}
	\bar{V}_{;\alpha \gamma} B^{\gamma \beta} f_{,\beta} &=& 
	\left( \hbar \Omega \right)^2 f_{,\alpha} .
\end{eqnarray}
The parameters $\omega$ and $\mu$ are Lagrange multipliers that
implement the condition of fixed angular momentum along the $x$ axis,
and fixed particle number, respectively.  In nuclear physics these are
usually called generalised cranking parameters.  The parameter
$\Lambda$ is an additional Lagrange multiplier for the collective
mode, forcing the system to stabilise at a point away from equilibrium.
The covariant derivative $V_{;\alpha \beta}$ is defined in the usual
way~\cite{NW99}.  The collective path is found by solving
Eqs.~(\ref{eq:force1}) and~(\ref{eq:localRPA1}) self-consistently,
i.e., we look for a path consisting of a series of points where the
lowest non-spurious eigenvector of the local RPA equations also
fulfils the force condition.  In the minimum of the potential the
spurious solutions decouples from the other collective and
non-collective solutions. In this paper we have chosen to ignore the
effects of the spurious admixtures to the RPA wave-functions, but
test calculations has showed these to be small. To limit the computational 
effort we use the method presented in Ref.~\cite{NW99,AW04} to reduce the size
of the RPA matrix. There it was shown that the RPA equation can be
solved with good accuracy by assuming that the RPA eigenvectors can be
described as a linear combination of a small number of state-dependent
one-body operators. The same basis set turned out to work well also at finite 
angular momentum.

We apply the LHA to the constrained pairing+quadrupole Hamiltonian as described 
in~\cite{BK68} with a constraint on particle numbers and angular momentum 
\begin{equation}
	\label{eq:HPQ2}
	H' = h_0 - \sum_\tau G_\tau P^\dagger_\tau P_\tau - \frac{\kappa}{2} 
	\sum_{M=-2}^2 Q^\dagger_M Q_M - \omega J_x - \sum_\tau \mu_\tau N_\tau,
\end{equation}
where $h_0$ is the spherical Nilsson Hamiltonian~\cite{NR95}. $Q_M$ and $P_\tau$ 
are the dimensionless quadrupole and pairing operators \cite{BK68}.
This Hamiltonian is treated in the Hartree-Bogoliubov approximation, 
within a  model space consisting of two major shells. We follow Ref.~\cite{BK68} 
and multiply all quadrupole matrix elements with a suppression factor. 

\begin{figure}%[htbp]
\centerline{\includegraphics[clip,width=7.5cm]{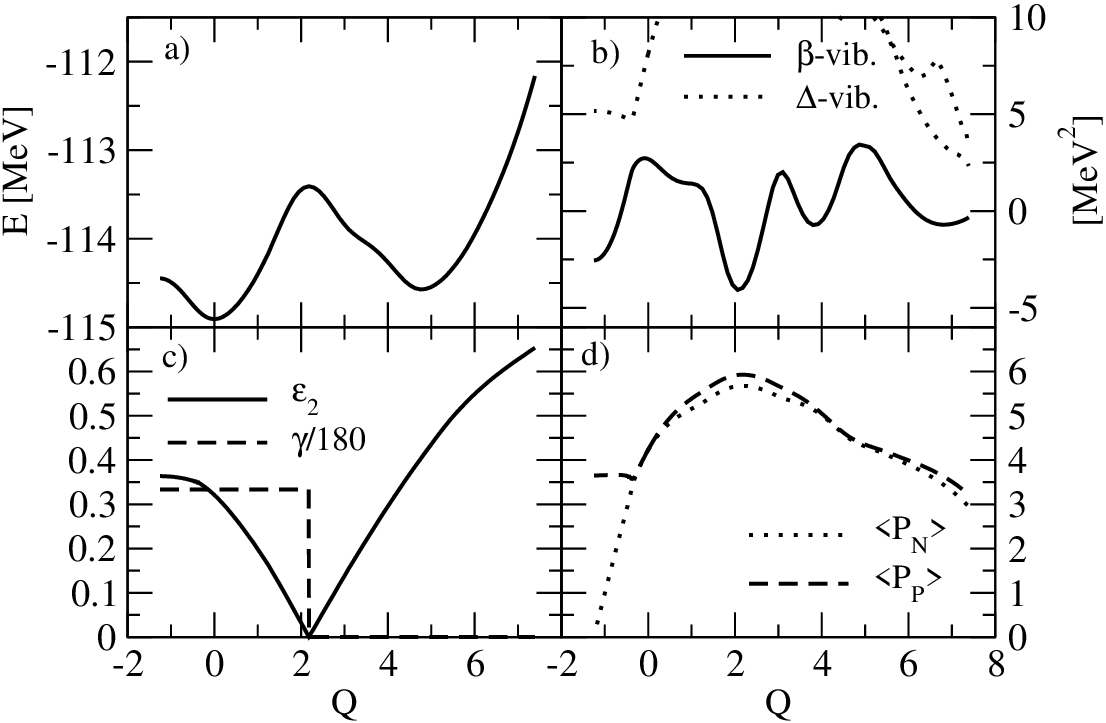}}
\caption{Large amplitude collective motion in $^{72}$Kr at $I=0$. 
  $Q$ is the collective coordinate. a) Energy along the collective path.  b)
  The square of the lowest RPA frequencies.  c) The deformation $\epsilon_2$
and the triaxiality $\gamma$. d) The dimensionless pairing
  operators $\left< P_\tau \right>$.  
}
  \label{fig:kr72I0}
\end{figure}
We start by examining the non-rotating states in $^{72}$Kr where a
prolate-oblate shape coexistence is established
experimentally~\cite{BM03}. We find a collective path going from the
oblate minimum over a spherical energy maximum into a prolate
secondary minimum and continuing towards lager deformation, see
Fig.~\ref{fig:kr72I0}. This is in contrast to the result recently found 
for $^{68}$Se~\cite{KN03} where the non-rotating prolate and oblate 
minima are connected via non-axial states. At a value of the collective coordinate
$Q\approx 7.5$ there is an avoided crossing between the lowest RPA
mode, which we are following, and two higher lying modes that are of
pairing vibrational character as can be seen in Fig.~\ref{fig:kr72I0}b. 
At this point our algorithm is no longer able to find a stable solution. 
This indicates that more than one collective coordinate should be used,
which at the moment is a very difficult calculation.
At large oblate deformation we see a collapse in neutron pairing
after an avoided crossing between the $\beta$-vibration we
are following and a pairing vibration.
As explained in detail in Ref.~\cite{AW04}, such points  are 
like the origin of radial coordinates for the collective motion, and it is not correct
to continue the calculation. Clearly this does not preclude other collective
coordinate becoming important here.

At finite angular momentum the collective path will no longer go
through the spherical state, since the rotational energy diverges for
such a state. Instead the oblate and prolate minima are connected
by a path consisting of non-axial states. Due to problems with
narrow level crossings we have to start in both the prolate and oblate minima
to find the complete collective path.
\begin{figure}%[htbp]
\centerline{\includegraphics[clip,width=7.5cm]{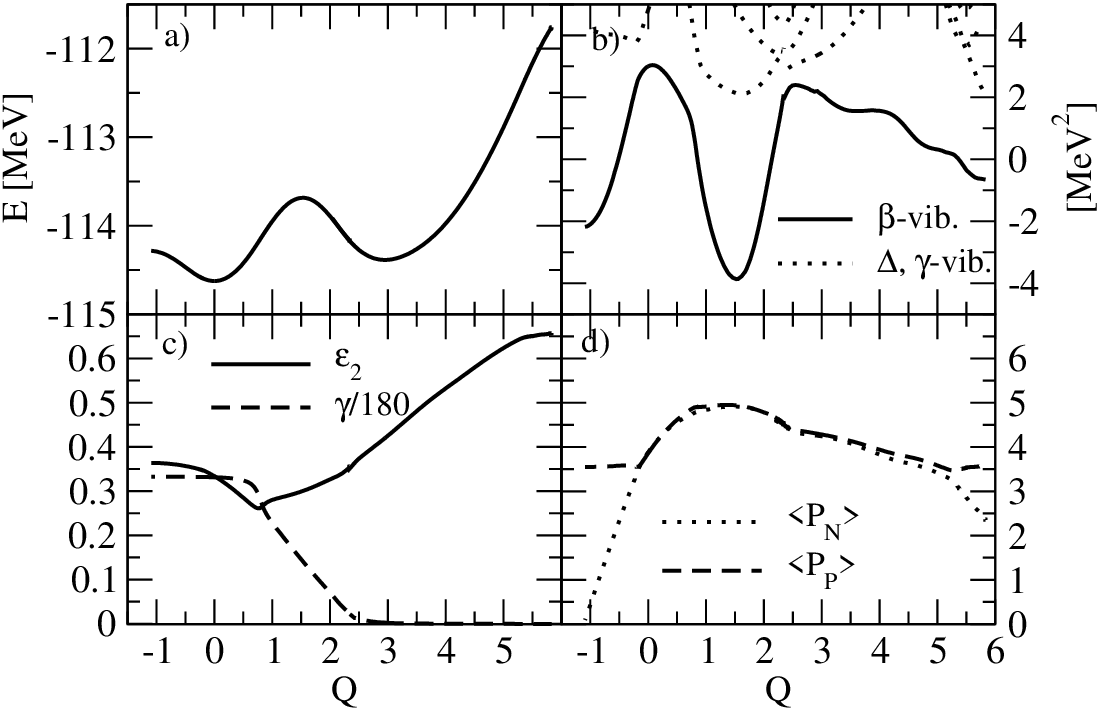}}
\caption{Large amplitude collective motion in $^{72}$Kr at $I=2$. 
  See Fig.~\ref{fig:kr72I0} for an explanation of the various figures.}
  \label{fig:kr72I2}
\end{figure}
In Fig.~\ref{fig:kr72I2} we see that if we start in the (almost)
oblate minimum and follow the collective path towards larger oblate
deformation the situation is very similar to the non-rotating case
discussed above: We first see a decrease in the the quadrupole
deformation, but after an avoided crossing with a pairing vibration
the collective path changes its nature and ends in
a neutron pair-field collapse. When we follow the collective path in
the other direction, towards smaller oblate deformation the system
goes through an avoided crossing with a $\gamma$-vibrational mode. The
path then turns into the triaxial plane and the $\epsilon_2$
deformation is almost constant, but with $\gamma$ decreasing quickly
from $60^\circ$ to $0^\circ$. At this point the collective path goes
through another avoided crossing this time with a
$\beta$-vibration. After the crossing the collective path follows an
almost prolate shape with increasing deformation. The number of
avoided crossing suggests that our description, even though it
captures much of the physics, is not fully quantitative.

\begin{figure}%[bhtp]
\centerline{\includegraphics[clip,width=7.5cm]{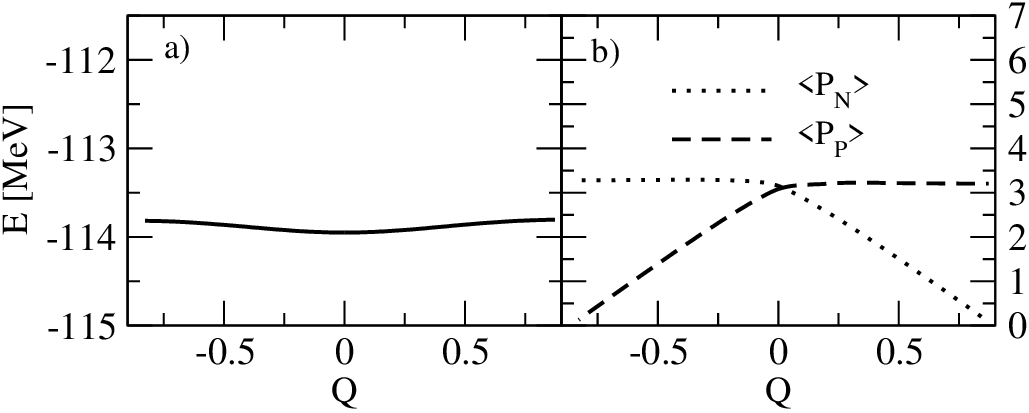}}
\caption{Large amplitude collective motion in $^{72}$Kr for $I=4$.
  $Q$ is the collective coordinate. a) Energy along the collective path.  b)
  The dimensionless pairing operators $\left< P_\tau \right>$.}
  \label{fig:kr72I4}
\end{figure}

In Fig.~\ref{fig:kr72I4} at angular momentum $I=4$ we see a very 
different character of the collective path.
The collective path is limited by the collapse of both the proton and
neutron pair-field. The points of zero neutron or proton pairing has
an excitation energy of less the $150$ keV.  When starting the
calculation in the prolate minimum we only find the collective path in
a limited region due to avoided crossings. This suggests that this
part of the collective dynamics is irrelevant. A related calculation
at zero gap, which hardly changes the minimum energy, but has
different fluctuations, suggests that there is no collective dynamics
around the $4^+$ state calculated here, and thus no shape mixing or
shape coexistence.

\begin{figure}%[bhtp]
  \centerline{\includegraphics[clip,width=7.5cm]{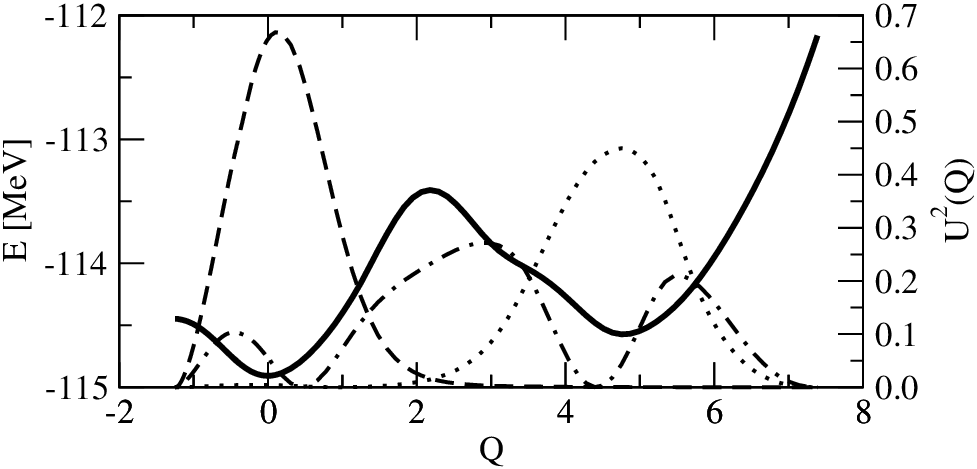}}
  \centerline{\includegraphics[clip,width=7.5cm]{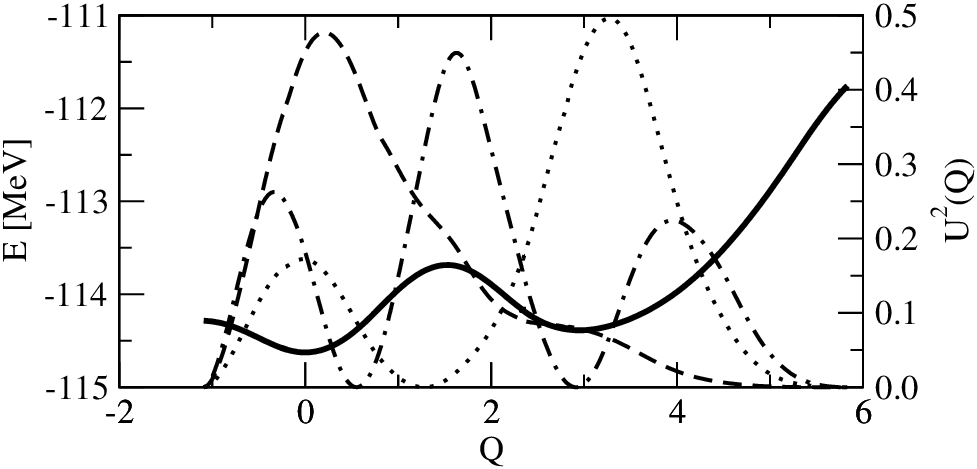}} 
%  \centerline{\includegraphics[clip,width=7cm]{kr72I43Dm.eps}} 
  \caption{The wave function for the large amplitude collective motion
  in $^{72}$Kr at $I=0$ and $2$ as a function of the collective
  coordinate $Q$. The thick solid line is the potential energy. The
  wave functions have their scale on the right side.} 
  \label{fig:kr723D}
\end{figure}

We want to solve the collective Hamiltonian along the collective path.
After having made a semi-classical approximation, which leads to a
classical Hamiltonian, we need to remember that we are studying a
quantum system. The standard technique to deal with this is to treat
the classical Hamiltonian as a quantum one, and to calculate the
eigenfunctions and energies. Details on how we do this can be found in
Ref.~\cite{AW04}. We have solved the collective Hamiltonian for the
$I=0$, $2$ and $4$ cases in $^{72}$Kr. We also need to calculate the 
proton and neutron pairing-rotational masses. They becomes zero at the 
point where $\left< P \right> \rightarrow 0$. In Fig.~\ref{fig:kr723D} 
we see that the ground state wave-function is concentrated in the
oblate minimum, but that it has a substantial spread
along the collective path and is skewed towards the spherical state
due to the collapse in the neutron pair-field. The first excited $I=0$
state has its major component in the prolate minimum with a small
component in the oblate minimum. The prolate peak in the wave-function
for the first excited state is broader and more symmetric than for the
ground state. The third $I=0$ state is approximate spherical but has
substantial oblate as well as prolate components. The prolate state
lies at a very low excitation energy of only $0.37$ MeV, while the
spherical state is somewhat higher in energy.
For $I=2$ the situation is similar to the case $I=0$. However, the
mixing of the prolate and oblate states is substantially stronger. The
long tail of the oblate peak stretches along the collective path into
the prolate minimum, and that there is a secondary oblate peak in the
prolate state.  Due to the wide peaks of the collective wave function
the expectation value of deformation in the collective states are
substantially different from those of the minimum potential energy
states, as can be seen in Table~\ref{tab:Edef}. Note that even though
the mean-field results show an almost axially symmetric solution at
$I=2$ ($\gamma < 1^\circ$) the collective state shows a substantial
$\gamma$-deformation.
\begin{table}%[htb] 
\caption{The deformation and excitation energy (in MeV) for the collective states.}
\begin{ruledtabular}
\begin{tabular}{r|cccccccc}
\label{tab:Edef}
$I_n$ & $0_1$ & $0_2$ & $0_3$ & $2_1$ & $2_2$ & $2_3$ & $4_1$ & $4_2$ \\ 
\hline
$\epsilon_2$ & 0.28 & 0.37 & 0.15 & 0.30 & 0.40 & 0.36 & 0.36 & 0.45 \\ 
$\gamma$ & 60.0 & 0.0 & 0.0 & 42.4 & 9.8 & 19.9 & 59.3 & 0.7 \\ 
$E_{ex}$ &  & 0.37 & 0.96 & 0.32 & 0.61 & 1.33 & 1.04 & 1.07 
\end{tabular}
\end{ruledtabular}
\end{table} 
For the $I=4$ state the situation is different. We can use the 
pairing collective paths, but these are
limited by the pairing collapse. Therefore we do not see any low lying
excited states and we see no indication of a  collective wave
function extended into the plane of non-axial deformation.
\begin{figure}
  \centerline{\includegraphics[clip,width=7.5cm]{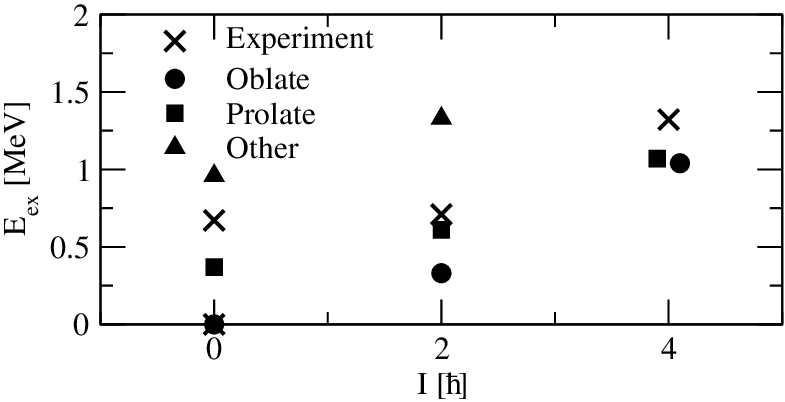}}
    \caption{Excitation energy of the collective states for $I=0$, $2$ and $4$ %
    as a function of angular momentum. Experimental data is taken from~\cite{BM03}.}
    \label{fig:kr72Edef}
\end{figure}

Figure~\ref{fig:kr72Edef} shows the collective excitation spectrum for
$I=0$, $2$ and $4$ in $^{72}$Kr as a function of the angular
momentum. We can see that the prolate and oblate states get close in
energy at finite angular momentum. We have also plotted the
experimental data~\cite{BM03} for the low spin states. The ground
state is thought to be oblate but the rotational band build on to of
it is thought to be prolate. The second experimental $I=0$ state is
interpreted as the prolate band-head. Our calculations support these
conclusions with the exception of the interpretation of the second
$0$-state. We find the prolate band head to be lower-laying in energy
then seen in experiment and we also see very little mixing of the
oblate and prolate states at $I=0$. An alternative interpretation
would be that the second $0$-state is a spherical state in our
calculation, since this state has a higher energy and a lager mixing
with the oblate state.

In summary, we have extended the method of calculating the
self-consistent collective path presented in~\cite{DK00,AW04} to
include constraints on angular momentum. We have been able to
determine the collective coordinate from the Hamiltonian without
having to assume a priori which are the relevant degrees of freedom.
The results confirm the importance of pairing collapsed states for the
collective path as suggested in Ref.~\cite{AW04}. We have also seen
the effect of rotation on the collective path: Without rotation the
path goes through a spherical saddle-point in contrast to the rotating
case where the two minima are connected via the triaxial plane, and as
we increase angular momentum the collective path disappears. The
changes would probably be less pronounced if we would have allowed for
more the one collective coordinate, which a calculation we intend to
do in the near future.  We have compared our calculations with
experimental data~\cite{BM03} and found a reasonable agreement.

%\section*{Acknowledgement}
This work was supported by the UK Engineering and Physical Sciences Research 
Council (EPSRC)

\end{document}